\documentstyle[twocolumn,aps,prl,epsf]{revtex}
\begin{document} 

\title{Revisiting the Theory of Finite Size Scaling in Disordered
Systems: \hskip 8cm $\nu$ Can Be Less Than 2/d}
\author{Ferenc P\'azm\'andi, Richard T. Scalettar and Gergely T. Zim\'anyi}
\address{Physics Department, University of California, Davis, CA 95616}
\address{\mbox{ }}

\address{\parbox{14cm}{\rm \mbox{ }\mbox{ }
For phase transitions in disordered systems, an exact theorem provides 
a bound on the finite size correlation length exponent: 
$\nu_{FS}\ge 2/d$. It is believed that the true critical exponent 
$\nu$ of a disorder induced phase transition satisfies the same bound. 
We argue that in disordered systems the standard averaging introduces
a noise, and a corresponding new diverging length scale, characterized 
by $\nu_{FS}=2/d$. This length scale, however, is independent of 
the system's own correlation length $\xi$. Therefore $\nu$ can be 
less than $2/d$. We illustrate these ideas on two exact examples, with
$\nu < 2/d$. We propose a new method of disorder averaging, which achieves
a remarkable noise reduction, and thus is able to capture the 
true exponents.}}
\address{\mbox{ }}
\address{\parbox{14cm}{\rm \mbox{ }\mbox{ }
}}
\address{\mbox{ }}
\maketitle

\narrowtext

Using a very general formulation, Ref.\cite{chayes} presented an exact
theorem, which puts constraints on the finite size correlation 
length exponent $\nu_{FS}$ of a large class of disordered systems:
$\nu_{FS}\ge 2/d$, where $d$ is the dimension. This relation is 
often referred to as the quantum Harris criterion\cite{harris}.
While many investigations found exponents in accordance with this bound,
there is an increasing number of results in contradiction with it.
In particular, in a model for charge density waves exact calculations
yielded $\nu=1/2$ below four dimensions\cite{narayan}, and
numerical studies on 2d disordered Bose-Hubbard models found
$\nu\simeq 0.7$ \cite{rieger}. Experimentally the Bose glass transition 
of helium in aerogel\cite{crowell}, and the localization transition in 
doped semiconductors\cite{rosenbaum} seem to violate this bound.
In this paper we argue that the $\nu_{FS} \ge 2/d$ constraint is
characteristic only to the method used to carry out the disorder 
average, and the true exponent $\nu$ is independent of this bound.

To start our considerations of random sytems, 
we chose the same type of disorder used by Ref.\cite{chayes}:
a binary distribution for, say, a disordered site energy.
Typically, physical quantities are calculated by averaging over
different disorder realizations. For calculational convenience, 
the standard method is analogous to the ``grand canonical" approach: 
impurities are put on each site with a given probability, $p$, and the 
averaging is carried out for all possible concentration of impurities 
and their configurations. 
An alternative method, which could be termed the
``canonical" approach, keeps the number of impurities fixed, and the average 
is taken only over the possible configurations of these impurities.
For infinite systems the two methods are equivalent.
The density fluctuations in the grand canonical method, however, introduce 
an extra noise. This noise vanishes in the infinite system, but 
it may alter the results of the finite size scaling. 
The ``canonical averaging" strongly reduces this noise by excluding
density fluctuations. 

We now argue that the bound obtained in Ref.\cite{chayes}
is only generated by the noise introduced by the ``grand canonical averaging".
Different choices, such as using ``canonical averaging", 
produce different bounds. 
The theorem of Ref.\cite{chayes} considers  a random system 
where a phase transition is induced by changing
the concentration $K$ of site (or bond) impurities. 
Let $Y$ be any event depending on disorder realizations
in a finite volume, with probability ${\cal P}(K)$.
This ${\cal P}(K)$ is calculated by {\it averaging} over
all disordered configurations, and selecting those compatible with $Y$. 
Averaging is performed in the ``grand canonical" way,
since fluctuations in the density of impurities are allowed. 
 From these premises the exact statement 
$|d {\cal P}(K)/d K|~\le~const.~\sqrt{N}$ 
follows, where $N$ is the system size. 
A closer look at the proof reveals that this result 
is derived solely from the concentration fluctuations 
of the impurities, which were {\it externally introduced in the 
averaging process} (see the last equation of the proof 
in Ref.\cite{chayes}). Thus the bound on $|d {\cal P}(K)/d K|$ does not relate
to the intrinsic properties of the system under investigation.
It only reflects the ``resolution" of the ``grand-canonical averaging". 
In other words, because of the presence of the density fluctuations,
the minimal resolvable change in $K$ is $dK \propto 1/{\sqrt N}$.
The probability ${\cal P}$ can change at most ${\cal O}(1)$, 
immediately explaining the above bound.

On the other hand, if one uses ``canonical averaging",
then the above inequality does not apply. 
For, in contrast to the previous case, the number of impurities
is now well defined. In the present binary example,
the resolvable change of $K$ is bounded only by its minimum allowed increment, 
$1/N$. Hence, $|d{\cal P}(K)/d K|~\le~N~.$
Along the lines of Ref.\cite{chayes}, the
inequality $\nu_{FS}\ge 1/d$ now follows.
As before, this inequality is characteristic
of the ``canonical averaging" only, and does not impose any restriction
on the true exponent $\nu$ of the physical system. 
The physical reason behind this is that both averaging procedures 
introduce a {\it new characteristic length scale}, which has the
potential to obscure the true correlation length of the physical system.

It is also important to note that the assumption of a binary disorder
plays a crucial role in deriving the above bounds. For continuous 
distributions they do not necessarily apply. To see this,
consider the following simple example, motivated by the quantum phase 
transition
between the so called Mott-Insulator and Bose-Glass phases, which
takes place in interacting bose systems with
site disorder. At this transition the renormalization group flows are
controlled by a fixed point with {\it zero} hopping strength\cite{us},
thus the system reduces to a collection of independent sites with random 
energies.
Let the distribution of the site energy $\epsilon\in [0,K]$ be 
\begin{equation}
P(\epsilon)=
{{\alpha+1}\over K^{\alpha+1}}\bigl(K-\epsilon\bigr)^\alpha,
\label{Pe}
\end{equation}
whith $\alpha>-1$. We generate $N$ independent $\epsilon_i$ 
($i=1,\dots,N$) from the above distribution. 
We define the finite-size event $Y$ to occur, when {\it all} $\epsilon_i$'s
are smaller then a given value $\mu\in (0,K]$.
We fix the value of $\mu$, and drive the transition by changing $K$.
As required by the theorem of \cite{chayes}, the probability ${\cal P}$
of $Y$ happening is finite at the critical value of the disorder, $K_c=\mu$.
It goes to zero exponentially with the system size $N$ for $K>K_c$.
Close to the transition, for $\delta=(K-K_c)/K_c\ll 1$, this
probability is
\begin{equation}
{\cal P}(N,\delta)\simeq e^{-N\delta^{\alpha+1}} ~~.
\label{MIBG}
\end{equation}
A characteristic length scale $\xi_f$ can be now defined as a function of
$\delta$. It is determined from the system size as $N_f=\xi_f^d$, 
where ${\cal P}(N_f,\delta)/{\cal P}(N_f,0) \sim 1/e$. 
Defining a critical exponent as $\xi_f \propto \delta^{-\nu_{FS}}$ one arrives
at $\nu_{FS}=(\alpha+1)/d$.
For $\alpha < 1$, $\nu_{FS}$ {\it is less than} $2/d$.
While we considered a concrete example, we emphasize that this result 
can be relevant for {\it any} transition driven by {\it local singularities} 
in the action.

Motivated by the above observations, we now attempt to
construct a {\it modified  finite size scaling procedure}, which does 
have the potential to access the true exponents of the system.
The centerpiece of our argument is the following observation.
If the distribution of the disorder is given in 
an analytic form, that uniquely determines $K_c$,
the critical value of the control parameter
for the infinite system. However, any {\it given} disorder realization 
in a {\it finite system} could have been generated from disorder 
distributions with a {\it range} of parameters, corresponding 
to a {\it range} of $K_c$ values. In other words, it is unclear
{\it which} infinite system's finite size realization did one simulate.
A distribution is characterized completely by its moments. 
Typically, $K_c$ is linked to some of these moments, for instance 
the dispersion. For a finite system of size $N$, this dispersion
is determined only with a relative uncertainty of 
${\cal O}(1/\sqrt{N})$. Therefore there is a range of distribution
parameters which are compatible with the specific realization, and 
thus could have generated it. This raises the problem, {\it which 
$K_c$ to use} in a finite size scaling analysis.

The standard procedure answers this question by assuming that one can 
use a single $K_c$ for all samples generated from the same distribution. 
However, the above argument suggests that the very same sample may 
be the realization of distributions with different parameters,
leading to an inherent noise in the procedure, similar to the above
considered binary examples. In order to avoid such a built-in noise,
we now propose a modified finite size scaling procedure for
disordered systems.
We suggest that for {\it each disorder realization}
one should identify the distribution, and in particular the 
critical parameter $K_{c}^r$, 
which it {\it most likely} corresponds to. In practice this
might be difficult, and we return to this question later.
For the moment, we only assume that it is possible to identify $K_{c}^r$.
We propose that the natural control parameter of the critical 
behaviour is  $\Delta = (K-K_{c}^r)/K_{c}^r$.
The act of averaging should then be performed for the samples 
with the same $\Delta$.
We propose to adopt the finite size scaling hypothesis for
the critical behaviour of a generic physical quantity $Q$,
\begin{equation}
\bar{Q}(L,\Delta) = L^{-y}q(L\Delta^{\nu}) ~~,
\label{Q}
\end{equation}
where $q(z)$ is a universal scaling function, and $y,\nu$ are the 
critical exponents for $Q$, and the true correlation length 
$\xi \propto \Delta^{-\nu}$. 
Note that some aspects of this proposition are already practiced
in numerical studies: sizeable noise-reduction is customarily reached by
adjusting the random variables {\it after} they are generated,
e.g. in order to keep their mean value constant. 

Next we assume the validity of Eq.\ref{Q}
and perform the standard finite size scaling,
to demonstrate how that procedure's inherent noise
can mask the true critical behaviour. 
Some of the key results of the analysis are: $i)$ we find that the
exponent of the intrinsic correlation length $\nu$ might be different 
from $\nu_{FS}$ appearing in the standard finite size scaling.
Therefore the theorem of Ref.\cite{chayes} does not provide 
constraints on the true exponent $\nu$. 
$ii)$ In particular, $\nu$ {\it can be} less than $2/d$. In this case
typically $\nu_{FS}=2/d$.

The standard finite size scaling procedure\cite{cardy} in disordered
systems calls for calculating a physical quantity, $Q$, such as
the critical susceptibility, for
different values of $N$ and $K$, the system size and
control parameter, each time performing the calculations for a number
of disorder realizations. Averaging over the disorder yields 
$\langle Q(K) \rangle$, and the critical coupling $K_c$ is then
identified for instance from a crossing pattern\cite{binder}.
Requiring the collapse of the data, when plotted as a function 
of $L^{1/\nu}\delta$, where $\delta=(K-K_c)/K_c$, determines the exponents.

To make contact between the standard scaling procedure and Eq.\ref{Q},
a relation between the unique $K_c$ and the fluctuating $K_c^r$ has
to be constructed. A simple representation of the inherent noise, 
or uncertainity, is to assume the validity of the central limit 
theorem for $K_c^r$
\begin{equation}
\Delta=\delta + \frac{D}{L^{d/2}} x ~~,
\label{Delta}
\end{equation}
where $x$ is a random variable with a distribution width of ${\cal O}(1)$.
Here $D$ measures the scatter in $K_{c}^{r}$, and $\delta$ is the distance
from the average critical point $K_c$. As we have seen, this is not necessarily
true for all systems, and we will return to the case when the 
fluctuations scale with a different power. 

The standard procedure neglects the fluctuations of $K_{c}^r$,
which is equivalent to averaging $\bar{Q}$ over the random 
variable $x$ of Eq.\ref{Delta}:
\begin{equation}
\langle Q \rangle = L^{-y} {\Bigg\langle} q \Biggl( D^{\nu}
L^{1-\frac{d\nu}{2}} \biggl( x+\frac{\delta L^{d/2}}{D}\biggr)^{\nu}
\Biggr) {\Bigg\rangle} ~~.
\label{ave}
\end{equation}
Here the $x$ average is denoted by $\langle\dots\rangle$,
corresponding to the standard averaging procedure, as opposed to ${\bar{Q}}$, 
the correlated averaging of the new procedure
in Eq.\ref{Q}. 

First we analyze the critical point itself, then we shall proceed
to extract the critical behaviour of the correlation length.
At $\delta =0$ the scaling form for $Q$ is
\begin{equation}
\langle Q \rangle = L^{-y} \Big\langle q \Bigl( D^{\nu}x^{\nu}
L^{1-\frac{d\nu}{2}}\Bigr)\Big\rangle ~~.
\label{L}
\end{equation}
For $\nu > 2/d$ the argument of the scaling function approaches zero
with increasing system size, and the $L$ dependence of the {\it averaged}
quantity $\langle Q (L)\rangle$ is characterized by the
{\it intrinsic} exponent $y$.
Here we use the customary assumption that the universal scaling 
function $q(z)$ approaches a finite value as $z\rightarrow 0 $. 

In the $\nu < 2/d$ case, however, the argument of $q(z)$ goes to
large values, probing deeply non-critical regions, even though
the system is assumed to be {\it at criticality}.
To highlight the consequences of this, we proceed with a
generic form for the asymptotic behaviour of the scaling function,
adopting $q(z) \propto z^{-\beta}$. From Eq.\ref{L}
$\langle Q \rangle \propto L^{-\gamma}$,
where $\gamma=y+\beta\bigl(1-d\nu/2\bigr)$.
Clearly the $L$ dependence of the averaged $\langle Q \rangle$ is 
{\it different} from the intrinsic value $y$.

Next we develop an understanding of the region in the proximity of the
critical point, i.e. the case of finite $\delta$.
Let us first focus on $\nu <2/d$. From Eq.\ref{ave} one identifies 
two scaling regions, governed by {\it two different} characteristic 
diverging length scales.

For large system sizes inevitably $D^\nu L^{1-d\nu/2}\gg 1$ , so the argument
of $q(z)$ again extends to large values. Utilizing
the previous asymptotic model form,  
\begin{equation}
\langle Q \rangle = L^{-\gamma} {\hat{q}} \bigl( \delta L^{d/2}\bigr) ~~.
\label{FS}
\end{equation}
from which a length scale can be identified, characterizing the
finite size scaling of $\langle Q \rangle$, averaged in the 
standard way. It diverges with an exponent $\nu_{FS} = 2/d$
{\it even though the true exponent} $\nu$ {\it is less than} $2/d$.
This result now demonstrates in general, what has been observed 
earlier for the binary example: the standard, or ``grand canonical''
averaging introduces a noise, which in turn generates a
new length scale and a corresponding new exponent into the analysis.

The other scaling region is reached when $\delta L^{d/2}/D \gg 1$.
In this limit
\begin{equation}
\langle Q \rangle = L^{-y} {q} \bigl( \delta^\nu L\bigr) ~~.
\label{nu}
\end{equation}
As is known, for large values of $\delta^\nu L$, the 
$\nu$ exponent is not accessible by finite size scaling\cite{cardy},
hence $\delta^\nu L$ should be kept around unity.
Therefore the determination of $\nu$ requires the study
of the region {\it away from the asymptotics}: large 
$\delta$ and small system sizes. For weak disorder ($D \ll 1$)
this window in fact might be wide enough for practical purposes.
To reiterate, however, studies concentrating on the 
asymptotic region are bound to see $\nu_{FS}=2/d$.

In the case of $\nu>2/d$ the standard procedure is capable
of accessing the true $\nu$: it can be obtained from 
$\langle Q \rangle$ by increasing the system size to the extent 
of $\delta L^{d/2}/D \gg 1$, but keeping
$\delta^\nu L\propto {\cal O}(1)$. This again implies
avoiding the ``non-scaling" region around $\delta=0$.
For strong disorder and small available system sizes,
one can end up again with large arguments of $q(z)$,
and consequently in the scaling regime described by $\nu_{FS}$
and $\gamma$ (Eq.\ref{FS}). There are several additional
crossover regimes in the parameter space, which can be studied 
based on Eq.\ref{ave}.

What happens if instead of the central-limit-theorem form, $L^{-d/2}$,
the fluctuations of $K_c^r$ are described by some other power law? For 
instance, on physical grounds, the fluctuations 
may scale with the correlation length exponent
\begin{equation}
\Delta=\delta + \frac{D}{L^{1/\nu}} x ~~.
\label{ok}
\end{equation}
Substituting this expression into Eq.\ref{Q} and averaging over $x$
shows that the standard procedure and our proposition give
the same result for the exponents, although the scaling function
changes due to the difference in averaging. 

Equation \ref{ok} can help incorporate the idea of
$K_c^r$ in the scaling. One can appreciate that from looking at a
finite sample, it is far from trivial to identify $K_c^r$,
belonging to the infinite system. A solution might be suggested
by recalling that for ordered classical magnets, the maximum of
the susceptibility of a finite size sample is shifted as:
$T_c(L)-T_c({\infty}) \propto L^{-1/\nu}$, just as in 
Eq.\ref{ok}. Scaling then can be performed in terms of $T_c(L)$, 
resulting in the correct exponents.
Even in the absence of knowledge of the $K_c^r$ of the corresponding infinite
system, one can still extract a $\tilde{K}_c^r$ from a specific 
feature of a critical quantity of the {\it finite size system}. 
Using this $\tilde{K}_c^r$ 
in our new scaling approach should provide the correct
exponent $\nu$, provided that $\tilde{K}_c^r - K_c^r \propto
L^{-1/\nu}$, a reasonable assumption.
 
We are thus left with the task of identifying $\tilde{K}_c^r$
of a finite system.
For many quantum systems at $T=0$ a reasonable proposition
for $\tilde{K}_{c}^r$ might be the value of $K$, where the gap 
to the first excitation vanishes or has a minimum. For classical systems 
$\tilde{K}_{c}^r$ may be identified where some critical susceptibility 
exhibits a maximum.

To demonstrate the above ideas, 
consider strongly interacting bosons in a random potential
at zero temperature. In Ref.\cite{us} renormalization flows were
generated by integrating out the sites with highest excitation 
energies. For infinite range hopping the renormalization group 
(RG) equations are {\it exact}. In particular, at the Mott-Insulator
to Superfluid transition weak disorder is irrelevant and $\nu=1/d$.
Because of the presence of an underlying RG one expects the
validity of finite-size scaling. 

\begin{figure}
\epsfxsize=3.0in
\epsfysize=2.25in
\epsffile{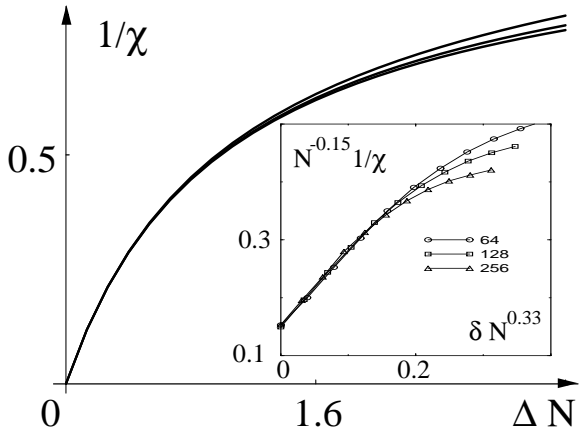}
\vskip 0.2cm
\caption{Scaling plot of the inverse susceptibility using the
novel and the standard (insert) averaging procedure for system
sizes N=64,128,256.}
\end{figure}

We carried out the finite-size scaling analysis of the 
average local susceptibility at weak
but finite disorder for system sizes $N=64,128,256$. 
First we used the standard averaging procedure (insert of Fig.1),
and we obtained $\nu_{FS}\simeq 3/d$ after averaging over 1024 realizations 
of a uniform disorder distribution of the random potential. The 
collapse of the curves for different system sizes is not perfect, and 
we expect that as $N$ increases, $\nu_{FS}\rightarrow 2/d$. Fig.1 shows
the same quantity scaled by using $\tilde{K}_c^r$ extracted from the
divergence of the susceptibility for each sample separately. 
The scaling is convincing, and yields the exact exponent
$\nu=1/d$. The exhibited curves were obtained by averaging over much fewer
samples than before, only 16, yet the scaling region extends by more than 
an order of magnitude further in terms of the scaling variable, $N\Delta$,
clearly demonstrating a very effective noise reduction. 
 
In some numerical studies, such as in Ref.\cite{rieger} a $\nu_{FS}<2/d$
has been reported, using the traditional averaging procedure. 
We would like to emphasize that this finding can be perfectly accomodated
in the present theory.
First, our analysis does {\it not} suggest that $\nu_{FS}$ {\it must} be 
greater or equal to $2/d$: this is only the most likely scenario. 
If, for instance, Eq.\ref{ok} describes the fluctuations of
$K_{c}^{r}$, then $\nu_{FS}=\nu$, and thus can be less than $2/d$. 
Apparently, this is the case in the example of the Mott-Insulator to Bose-Glass
transition in Eq.\ref{MIBG}. Second, as emphasized after Eq.\ref{nu}, 
if the fluctuations of $K_c^{r}$ are small and the sample size is not too big,
then the true $\nu$ can and will be observed in finite size scaling.
Finally, this theory is {\it not addressing} the problems associated
with distributions with long power-law tails \cite{dfisher}, or multicritical
fixed points \cite{singh}. After averaging, we still expect a gaussian
distribution, with exponential tails.

Now we would like to reflect on the Harris criterion \cite{harris}.
An insightful derivation imagines dividing the sample to blocks of size $\xi$.
For $\nu<2/d$, the fluctuations of the local ``$T_{c}$'s" of the blocks are 
bigger than the distance from the true $T_c$ and it is concluded that $\nu$
cannot be smaller than $2/d$. In our framework this negative result only 
means that the blocking procedure ceases to be a valid approach to
the infinite system, and says nothing about the value of the 
inherent exponents. An RG based investigation of this problem will 
be given elsewhere.

In sum, we reinvestigated the theory of finite size scaling in 
disordered systems. We found that the standard averaging procedure
introduces a new diverging length scale into the problem, therefore the finite 
size scaling exponent $\nu_{FS}$ may be unrelated to the inherent $\nu$
of the true correlation length. In particular, we constructed two
examples explicitly, where {\it exact} calculations proved that
the inherent $\nu$ is less than $2/d$. We proposed an alternative
method, which achieves a remarkable noise reduction, and therefore
is capable of accessing the true exponents of the physical problem.

We acknowledge useful discussions with
L. Chayes, T. Giamarchi, A. Otterlo, H. Rieger, 
S. Sachdev, R. Singh, P. Weichman, and P. Young. 
This research was supported by NSF DMR-95-28535.

\end{document}